\documentclass{article}
\usepackage{graphicx}

\newcommand{\figlab}[1]{\label{fig:#1}}

\newcommand{\figref}[1]{\ref{fig:#1}}
\newcommand{\eqref}[1]{(\ref{eq:#1})}

{\catcode`\@=11
\gdef\setft#1#2#3{%
\def\@oddfoot{
{\setbox0=\hbox{#1}
\setbox1=\hbox{#3}
\ifdim\wd0>\wd1
\dimen0=\wd0
\box0\hfil#2\hfil\hbox to\dimen0{\hfil\hfil\box1}
\else \dimen0=\wd1
\hbox to\dimen0{\box0\hfil }\hfil#2\hfil\box1 \fi
}}} }

\def\complaint#1{}
\def\withcomplaints{
\newcounter{mycomplaints}
\def\complaint##1{\refstepcounter{mycomplaints}%
\ifhmode%
\unskip%
{\dimen1=\baselineskip \divide\dimen1 by 2 %
\raise\dimen1\llap{\tiny -\themycomplaints-}}\fi%
\marginpar{\tiny [\themycomplaints]: ##1}}%
}

\usepackage{amssymb}

\def\a{{\alpha}}
\def\b{{\beta}}
\def\e{{\epsilon}}
\def\G{{\Gamma}}

\title{Computational Geometry Column 38}
\author{%
Joseph O'Rourke\thanks{
Dept. of Computer Science, Smith Col\-lege, North\-ampton, 
MA 01063, USA.
\-orourke@cs.\-smith\-.edu.
Supported by NSF Grant CCR-9731804.
}
}
\date{}
\begin{document}
\maketitle
\pagestyle{empty}
\thispagestyle{empty}

\begin{abstract}
Recent results on curve reconstruction are described.
\end{abstract}

Reconstruction of a curve from sample points
(``connect-the-dots'') is an important problem
studied now for twenty years.
Early efforts, primarily by researchers in computer vision,
pattern recognition, and computational morphology,
relied on ad hoc heuristics (e.g., my own~\cite{obw-cdnh-87}).
The heuristics were placed on a firmer footing with
$\a$-shapes~\cite{eks-sspp-83} and $\b$-skeletons~\cite{kr-fcm-85}
and other structures,
whose underlying proximity graphs
were later shown to support
accurate reconstruction from uniformly sampled
curves~\cite{fg-cmc-95,a-rrsru-98,bb-srmua-97}.  
User selection of the $\a$ or $\b$
parameter is still necessary.

A breakthrough was achieved by Amenta, Bern, and 
Eppstein~\cite{abe-cbscc-98},
who designed two algorithms that 
guarantee correct reconstruction of smooth closed curves even with
(sufficiently dense) 
$\underline{\hbox{non}}$uniform 
samples,
and which lift the burden of selecting a parameter.
One of their algorithms computes what they call the {\em crust\/},
a subgraph of the complete graph on the sample points that
coincides with the correct polygonal curve under the right conditions.
One of the novelties of their approach is to define the
sample density to increase on exactly those portions of the curve $\G$
where more points are needed for reconstruction.
They demand that for every point $x \in \G$, there is a sample
point $p$ such that $|xp| < \e \mu(x)$,
where $\mu(x)$ is the distance from $x$ to the medial axis/skeleton.
This distance is small 
wherever two sections of the curve are close (in the vicinity of
a sharp turn, or a narrow neck), for such sections are
separated there by a branch of the medial axis.
They can guarantee reconstruction for all $\e \lesssim \frac{1}{4}$.
Fig.~\figref{J}(a) illustrates a reconstruction using their algorithm,
and (b) shows the crust when the sample density is below their
threshold.\footnote{
	It should be noted that some applications call for multiple
components, or nodes of degree~$> 2$, as in (b) of the figure.}
%%%%%%%%%%%%%%%%%%%%%%%%%%%%%%%%%Figure Begin
\begin{figure}[htbp]
\centering
\includegraphics[width=0.75\textwidth]{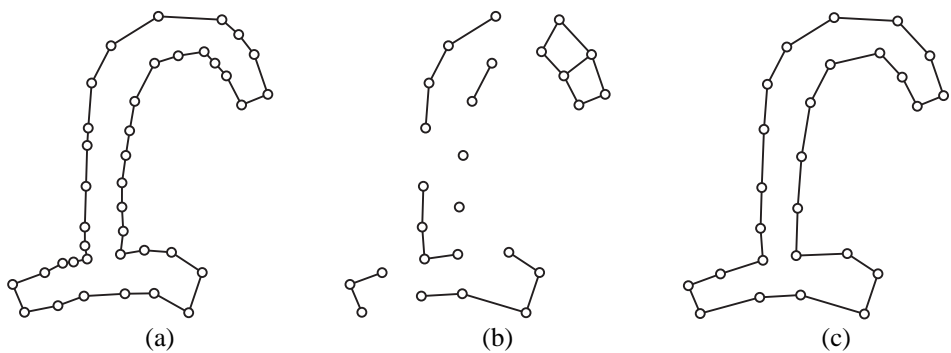}
\caption{
(a) \protect\cite{abe-cbscc-98} algorithm on a densely sampled curve.
(b) The crust~\protect\cite{abe-cbscc-98}, with $12$ fewer sample points.
(c) TSP algorithm on sparse set.
Computations performed at 
{\tt http://}
{\tt review.mpi-sb.mpg.de:81/Curve-Reconstruction/}
\protect\cite{amns-ecr-00}.
}
\figlab{J}
\end{figure}
%%%%%%%%%%%%%%%%%%%%%%%%%%%%%%%%%Figure End

Their work was followed by a flurry of improvements and extensions:
a computational improvement~\cite{g-cac-99},
a simpler ``nearest-neighbor crust'' that raises the density threshold to
$\e = \frac{1}{3}$~\cite{dk-spacr-99}, and
an extention to curves with endpoints~\cite{dmr-crcdgr-99}.
Another line of investigation was opened by 
Giesen~\cite{g-crtspmtl-99},
who proved that the TSP tour reconstructs the curve
for uniformly sampled nonsmooth curves
(and that no larger class of connected curves can be 
correctly reconstructed within the
Delaunay subgraph.)
This was quickly extended by Althaus and Mehlhorn
to nonuniform samples~\cite{am-tspbcrpt-00},
who in addition established that the TSP instance can be solved
in polynomial time.

Althaus et al.~\cite{amns-ecr-00}
have now implemented all the major curve reconstruction algorithms
using LEDA~\cite{mn-lpcgc-99},
and made them available for interactive comparison on the
Web.
Their experiments show that the TSP algorithm 
is both the most time-intensive computation
(13 times the fastest competitor, \cite{dk-spacr-99}), 
but also the most robust for sparsely sampled curves,\footnote{
	Or at least for certain classes of curves,
	with uniform random sampling.}
as indicated by Fig.~\figref{J}(c).

The next frontier in provable
reconstruction is reconstruction of 2D surfaces embedded in 3D.
See~\cite{abk-nvbsr-98,ab-srvf-99,ac-opdfh3dsr-99,dl-srs-99} for a start.

\subsubsection*{Acknowledgements}
I thank
E. Althaus,
N. Amenta,
M. Bern,
T. Dey,
D. Eppstein,
J. Giesen,
and
K. Mehlhorn for their comments.

\bibliographystyle{alpha}
\bibliography{/home1/orourke/bib/geom/geom}

\begin{thebibliography}{AMNS00}

\bibitem[AB99]{ab-srvf-99}
N.~Amenta and M.~Bern.
\newblock Surface reconstruction by {Voronoi} filtering.
\newblock {\em Discrete Comput. Geom.}, 22(4):481--504, 1999.

\bibitem[ABE98]{abe-cbscc-98}
N.~Amenta, M.~Bern, and D.~Eppstein.
\newblock The crust and the $\beta$-skeleton: {C}ombinatorial curve
  reconstruction.
\newblock {\em Graph. Models Image Process.}, pages 125--135, 1998.

\bibitem[ABK98]{abk-nvbsr-98}
N.~Amenta, M.~Bern, and M.~Kamvysselis.
\newblock A new {Voronoi}-based surface reconstruction algorithm.
\newblock In {\em Proc. SIGGRAPH '98}, Computer Graphics Proceedings, Annual
  Conference Series, pages 415--412, July 1998.

\bibitem[AC99]{ac-opdfh3dsr-99}
N.~Amenta and S.~Choi.
\newblock One-pass {Delaunay} filtering for homeomorphic {3D} surface
  reconstruction.
\newblock Manuscript, 1999.

\bibitem[AM00]{am-tspbcrpt-00}
E.~Althaus and K.~Mehlhorn.
\newblock Polynomial time {TSP}-based curve reconstruction.
\newblock In {\em Proc. 11th ACM-SIAM Sympos. Discrete Algorithms}, pages
  686--695, January 2000.

\bibitem[AMNS00]{amns-ecr-00}
E.~Althaus, K.~Mehlhorn, S.~N{\"a}her, and S.~Schirra.
\newblock Experiments on curve reconstruction.
\newblock In {\em Proc. 2nd Workshop Algorithm Eng. Exper.}, pages 103--114,
  January 2000.

\bibitem[Att98]{a-rrsru-98}
D.~Attali.
\newblock $r$-regular shape reconstruction from unorganized points.
\newblock {\em Comput. Geom. Theory Appl.}, 10:239--247, 1998.

\bibitem[BB97]{bb-srmua-97}
F. Bernardini and C.~L. Bajaj.
\newblock Sampling and reconstructing manifolds using alpha--shapes.
\newblock In {\em Proc. 9th Canad. Conf. Comput. Geom.}, pages 193--198, 1997.

\bibitem[FMG95]{fg-cmc-95}
L.~H. de~Figueiredo and J.~de~Miranda~Gomes.
\newblock Computational morphology of curves.
\newblock {\em Visual Comput.}, 11:105--112, 1995.

\bibitem[DK99]{dk-spacr-99}
T.~K. Dey and P.~Kumar.
\newblock A simple provable algorithm for curve reconstruction.
\newblock In {\em Proc. 10th ACM-SIAM Sympos. Discrete Algorithms}, pages
  893--894, January 1999.

\bibitem[DL99]{dl-srs-99}
T.~K. Dey and N.~Leekha.
\newblock Surface reconstruction simplified.
\newblock Manuscript, December 1999.

\bibitem[DMR99]{dmr-crcdgr-99}
T.~K. Dey, K.~Mehlhorn, and E.~A. Ramos.
\newblock Curve reconstruction: {C}onnecting dots with good reason.
\newblock In {\em Proc. 15th Annu. ACM Sympos. Comput. Geom.}, pages 197--206,
  1999.

\bibitem[EKS83]{eks-sspp-83}
H.~Edelsbrunner, D.~G. Kirkpatrick, and R.~Seidel.
\newblock On the shape of a set of points in the plane.
\newblock {\em IEEE Trans. Inform. Theory}, IT-29:551--559, 1983.

\bibitem[Gie99]{g-crtspmtl-99}
J.~Giesen.
\newblock Curve reconstruction, the {TSP}, and {Menger}'s theorem on length.
\newblock In {\em Proc. 15th Annu. ACM Sympos. Comput. Geom.}, pages 207--216,
  1999.

\bibitem[Gol99]{g-cac-99}
C.~Gold.
\newblock Crust and anti-crust: {A} one-step boundary and skeleton extraction
  algorithm.
\newblock In {\em Proc. 15th Annu. ACM Sympos. Comput. Geom.}, pages 189--196,
  1999.

\bibitem[KR85]{kr-fcm-85}
D.~G. Kirkpatrick and J.~D. Radke.
\newblock A framework for computational morphology.
\newblock In G.~T. Toussaint, editor, {\em Computational Geometry}, pages
  217--248. North-Holland, Amsterdam, Netherlands, 1985.

\bibitem[MN99]{mn-lpcgc-99}
K.~Mehlhorn and S.~N{\"a}her.
\newblock {\em {LEDA}: A Platform for Combinatorial and Geometric Computing}.
\newblock Cambridge University Press, Cambridge, U.K., 1999.

\bibitem[OBW87]{obw-cdnh-87}
J.~O'Rourke, H.~Booth, and R.~Washington.
\newblock Connect-the-dots: {A} new heuristic.
\newblock {\em Comput. Vision Graph. Image Process.}, 39:258--266, 1987.

\end{thebibliography}
\end{document}